\documentclass[floats,floatfix,showpacs,amssymb,prd,superscriptaddress,nofootinbib,aps,reprint]{revtex4-2}

\usepackage[final]{hyperref} 
\hypersetup{
    colorlinks=true,       
    linkcolor=blue,        
    citecolor=blue,        
    filecolor=magenta,     
    urlcolor=blue         
}

\usepackage{graphicx}
\usepackage{dcolumn}
\usepackage{bm}
\usepackage{amsmath,amsbsy}
\usepackage{comment}
\usepackage{yhmath}

\usepackage{color}

\def\vev#1{{\left\langle #1 \right \rangle}}

\newcommand{\bfx}{{\bf x}}
\newcommand{\bfk}{{\bf k}}
\newcommand{\bfkp}{{{\bf k}_\perp}}

\newcommand{\bfK}{{\bf K}}
\newcommand{\bfKp}{{{\bf K}_\perp}}
\newcommand{\bfKpp}{{{\bf K}'_\perp}}

\newcommand{\dd}{\text{d}}
\newcommand{\ee}{\text{e}}

\newcommand{\tot}{\text{tot}}
\newcommand{\tmax}{\text{max}}
\newcommand{\tmin}{\text{min}}
\newcommand{\Mpc}{\text{Mpc}}

\begin{document}


\title{Parity-breaking galaxy 4-point function from lensing by chiral gravitational waves}

\author{Keisuke Inomata}
\email{inomata@jhu.edu}
\affiliation{William H.\ Miller III Department of Physics \& Astronomy, Johns Hopkins University, 3400 N.\ Charles St., Baltimore, MD 21218, USA}

\author{Leah Jenks}
\email{ljenks@uchicago.edu}
\affiliation{Kavli Institute for Cosmological Physics, University of Chicago, 5640 South Ellis Ave., Chicago, IL 60637}

\author{Marc Kamionkowski}
\email{kamion@jhu.edu}
\affiliation{William H.\ Miller III Department of Physics \& Astronomy, Johns Hopkins University, 3400 N.\ Charles St., Baltimore, MD 21218, USA}

\date{\today}

\begin{abstract}
Recent searches for parity breaking in the galaxy four-point correlation function, as well as the prospects for greatly improved sensitivity to parity breaking in forthcoming surveys, motivate the search for physical mechanisms that could produce such a signal.
Here we show that a parity-violating galaxy four-point correlation function may be induced by lensing by a chiral gravitational-wave background.  We estimate the amplitude of a signal that would be detectable with a current galaxy survey, taking into account constraints to the primordial gravitational-wave-background amplitude.  We find that this mechanism is unlikely to produce a signal large enough to be seen with a galaxy survey but note that it may come within reach with future 21cm observations.
\end{abstract}

\maketitle


\section{Introduction}

Parity is a fundamental symmetry arising in physics, but we know from experiment and observation that our Universe is not perfectly parity invariant. Parity is notably violated in the Standard Model of particle physics via the weak force \cite{Lee:1956qn, Wu:1957my}, and some amount of parity violation in the early Universe is also necessary in order to produce the present day matter-antimatter asymmetry. Beyond these known sources, recent observations have suggested that signatures of parity violation may also be present in cosmological data. One arena of interest is in large-scale structure data. In particular, recent analysis of galaxy survey data from the BOSS survey has indicated evidence for a parity-breaking four-point correlation function (4PCF) \cite{Hou:2022wfj,Philcox:2022hkh}, which is the lowest order N-point correlation function for scalar quantities encoding parity information.  The possibility of parity violation in the galaxy distribution is made perhaps more interesting given a reported preference in Planck 2018 CMB polarization data for a nonzero value of the cosmic birefringence angle \cite{Minami:2020odp}, which, though not necessarily of the same origin, also hints at cosmological parity-violating physics \cite{Lue:1998mq}.

The possibility to seek parity violation in galaxy clustering was suggested briefly in Refs.~\cite{Jeong:2012df,Masui:2017fzw}, and precise algorithms to carry out such searches were developed in Ref.~\cite{Cahn:2021ltp}, capitalizing upon novel techniques \cite{Slepian:2015qza,Cahn:2020axu,Philcox:2021bwo,Hou:2021ncj} for the more general 4PCF.  The implementation with BOSS data and evidence for parity breaking in Refs.~\cite{Hou:2022wfj,Philcox:2022hkh} has motivated the study of physical models that could induce such signals. 
Though the observational evidence for parity violation in the galaxy distribution has since been questioned \cite{Philcox:2024mmz,Krolewski:2024paz}, the prospects for greatly improved sensitivities afforded by forthcoming galaxy surveys make a continued investigation of models warranted.

Most of the ideas for a parity-breaking 4PCF involve coupling of the inflaton to a vector or tensor field in the early Universe \cite{Jeong:2012df,Shiraishi:2016mok,Bartolo:2015dga,Creque-Sarbinowski:2023wmb,Philcox:2022hkh,Cabass:2022oap,Cabass:2022rhr}.  A 4PCF in the primordial curvature perturbation may in this case be mediated by the exchange of one of these vector or tensor particles.  If these interactions are chiral, then the 4PCF becomes parity-breaking.  The galaxy 4PCF then inherits this parity-breaking 4PCF. Phenomenological approaches different from Refs.~\cite{Hou:2022wfj,Philcox:2022hkh} have also been studied~\cite{Jeong:2012df,Jamieson:2024mau}.

In this paper, we explore the possibility that a parity-breaking 4PCF in the galaxy distribution can arise from lensing by gravitational waves (GWs). A galaxy 4PCF is induced by gravitational lensing, but if the GWs doing the lensing are chiral, the 4PCF will be (as we show below) parity-breaking as well. The idea is similar to the early-universe scenarios, except that here the parity breaking is mediated by the effects of the GW background at late times, rather than during inflation and the onset of structure formation. In this work, we calculate the parity-breaking contribution to the 4PCF induced by chiral GWs, estimate the detectability of such a signal, and briefly discuss scenarios that would provide a chiral GW background.
We note that, throughout this work, we focus on the lensing effects on the galaxy distribution on the (approximate) two-dimensional (2D) plane.
In this sense, the signals we discuss do not explain the parity-breaking signals reported in Refs.~\cite{Hou:2022wfj,Philcox:2022hkh}, which are based on the observation of the three-dimensional (3D) tetrahedron configurations of the galaxies.
Although we focus on the galaxy distribution on the 2D plane, the total observation system (2D plane + observer) is 3D, which enables us to discuss the parity violation in the lensing signals.

Our paper is organized as follows:  In Section \ref{sec:4pcffromlensing}, we show how lensing by chiral GWs induces a parity-breaking 4PCF. In Section \ref{sec:detectability}, we describe the estimators that can be used to seek this parity-breaking signal and estimate the smallest signal detectable by a current or forthcoming galaxy survey.  In Section \ref{sec:models}, we give a brief overview of the types of models that could lead to a chiral GW background and thus a parity-breaking 4PCF.  We conclude in Section \ref{sec:conclusion}. 
Throughout the paper, we employ natural units such that $c =\hbar = 1$, and Latin indices $i,j,k$ indicate spatial indices.

\section{4-point correlation function from lensing by GWs}
\label{sec:4pcffromlensing}

In this Section, we determine the signal of the 4PCF from lensing by GWs. We first calculate the 4PCF in terms of a general lensing power spectrum in Section~\ref{sec:4PCF-general}, then specify to such a power spectrum produced by GWs in Section~\ref{sec:4PCF-GW}.

\subsection{The 4-point correlation function
}
\label{sec:4PCF-general}
We first calculate the contribution to the 4PCF from lensing. We proceed with the simplest possible calculation that illustrates the relevant physics.  Consider a survey of a cubic volume $V$ in the Universe of dimensions $s$ (where $V = s^3$) viewed along the $z$ axis with $x$ and $y$ as the transverse directions. 
This setup with a finite volume leads to discretized Fourier modes and we take this setup to help for the comparison with real observation data.
We assume that the comoving distance $r$ to the survey volume is large compared with $s$.  We can thus neglect the effects of redshift evolution, often called the light-cone effect~\cite{Matsubara:1997zj}.  We also neglect redshift-space distortions, as they are not relevant for determining the parity dependence of the 4PCF \cite{Hou:2022wfj}.  
This setup approximates the 3D observation volume as the 2D observation plane perpendicular to the light-of-sight, which means that the parity-breaking signals in the 3D distribution of galaxies are erased in this setup. Instead, this setup focuses on the parity-breaking signals caused by the lensing effects between the 2D plane and the observer.\footnote{We can also understand this with the time reversal transformation. The crucial point is that the parity breaking on the 2D celestial surface has the specific handedness that we observe because of the propagation of photons \emph{forward in time} through a chiral GW background.  The chiral GW background is odd in the time-reversal transformation, and the propagation of photons forward in time explicitly chooses a time direction.  Thus, what we are seeing is not an observer- or direction-dependent effect.  Any observer anywhere in this Universe, looking in any direction, would see the same parity breaking on their 2D celestial sphere.
}
A position in this volume is denoted by a vector $\bfx=(x,y,z)\equiv (\bfx_\perp,z)$, where $\bfx_\perp$ is the position in the transverse direction.  The fractional galaxy number density perturbation at $\bfx$ is $\delta_{g,0}(\bfx)$.  The (unlensed) galaxy two-point correlation function, $\xi(r) \equiv \vev{\delta_{g,0}(\bfx)\delta_{g,0}(\bfx+{\bf r})}$, is statistically isotropic (i.e., a function only of the separation $r$), and here we assume the galaxy distribution to be Gaussian.

Since we are assuming $r \gg s$, every galaxy experiences the same deflection due to lensing by GWs along the line of sight.  Gravitational lensing implies that a galaxy at position $\bfx_\perp$ on the sky is observed to be at $\bfx_\perp +\delta \bfx$, where $\delta \bfx$ is the deflection (in the $x$-$y$ plane) due to lensing.  The deflection can most generally be written as \cite{Stebbins:1996wx, Kamionkowski:1997mp}
\begin{equation}
    (\delta\bfx)_i = \partial_i \phi(\bfx_\perp) + \epsilon_{ij} \partial_j \omega(\bfx_\perp),
\end{equation}
in terms of a scalar $\phi(\bfx_\perp)$ and pseudoscalar $\omega(\bfx_\perp)$.
Note that the subscript index $i$ denotes the two-dimensional deflection space.
The fractional perturbation in galaxy number density observed at some position $\bfx$ is then
\begin{equation}
\delta_g(\bfx) = \delta_{g,0}(\bfx+ \delta\bfx) \simeq \delta_{g,0}(\bfx) + (\delta\bfx) \cdot \nabla \delta_{g,0}(\bfx),
\label{eq:delta_g_x}
\end{equation}
where we note again $\delta_{g,0}(\bfx)$ is the perturbation in the absence of lensing.

\begin{figure}
        \centering \includegraphics[width=1.0\columnwidth]{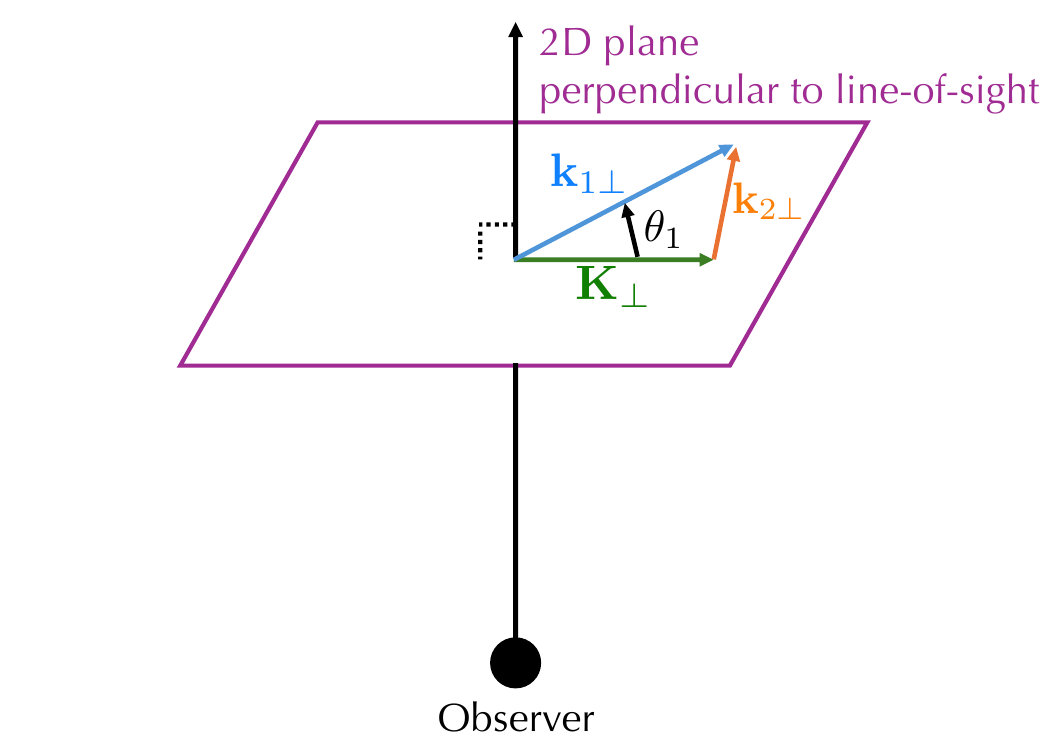}
        \caption{ 
        The relation of the vectors in Eq.~(\ref{eq:n_2pt}).
        }
        \label{fig:vecs}
\end{figure}

We now move to Fourier space where each wave vector can be written as $\bfk=(k_x,k_y,k_z) \equiv (\bfkp,k_z)$.  The Fourier amplitudes for the observed galaxy density are then,
\begin{widetext}
\begin{align}
    \delta_g(\bfk_\perp, k_z) &= \delta_{g,0}(\bfk_\perp, k_z) + S^{-1}\sum_{\bfKp} \left[ - \bfKp\cdot(\bfkp-\bfKp) \phi(\bfKp)  + \bfKp\times(\bfkp-\bfKp) \omega(\bfKp)\right] \delta_{g,0}(\bfkp-\bfKp,k_z) \nonumber \\
    &=  \delta_{g,0}(\mathbf k_\perp,k_z) + S^{-1} \sum_{\bfKp} K_\perp |\mathbf k_\perp - \bfKp| \left[-\cos \theta\, \phi(\bfKp) + \sin \theta\, \omega(\bfKp) \right] \delta_{g,0}(\mathbf k_\perp - \bfKp,k_z),
    \label{eq:delta_g_fourier}
\end{align}
where $S = s^2$ is the observed area in the 2D celestial space and $\theta$ is defined as the angle between $\bfKp$ and $\mathbf k_\perp - \bfKp$ with $\bfKp$ being the $x$-axis and the line-of-sight direction being the $z$-axis. See Appendix for the derivation of this expression.

We define the power spectrum for the unlensed perturbations through,
\begin{align}
    \vev{\delta_{g,0}(\mathbf k_{1\perp},k_{1z}) \delta^*_{g,0}(\mathbf k_{2\perp},k_{2z})} &= V\delta_{\mathbf k_{1}, \mathbf k_{2}} P_g(k_{1}),
    \label{eq:ps_3d}
\end{align}
where $\delta_{\bfk_1, \bfk_2}$ is the Kronecker delta.
Note that we take this normalization to make our power spectrum consistent with the power spectrum in infinite observation volume. For instance, $V \delta _{\bfk_1,\bfk_2} \to (2\pi)^3\delta_D(\bfk_1 - \bfk_2)$ in the infinite observation volume, where $\delta_D$ is the Dirac delta function.
The Fourier-space two-point correlation function then becomes 
\begin{align}
    \vev{\delta_g(\mathbf k_{1\perp},k_{1z}) \delta^{*}_g(\mathbf k_{2\perp},k_{2z})} =& \delta_{\mathbf k_1, \mathbf k_2} V P_g(k_1) - \frac{V}{S} \sum_{\mathbf K_\perp} \delta_{k_{1z},k_{2z}}\delta_{\mathbf k_{1\perp} - \mathbf k_{2\perp}, \mathbf K_\perp} K_\perp \nonumber \\
    & \times  \left\{
    -\phi (\mathbf K_\perp) \left[ k_{1\perp} P_g(k_1) \cos \theta_{1} + k_{2\perp} P_g(k_2) \cos \theta_{2} \right] \right. \nonumber \\
    & \left. \qquad  + \omega (\mathbf K_\perp) \left[ k_{1\perp} P_g(k_1) \sin \theta_{1} - k_{2\perp} P_g(k_2) \sin \theta_{2} \right] \right\},
    \label{eq:n_2pt}
\end{align}
where $\theta_{1}$ ($\theta_{2}$) are the angle between $\mathbf k_{1\perp}$ ($-\mathbf k_{2\perp}$) and $\mathbf K_\perp = \mathbf k_{1\perp} -\mathbf k_{2\perp}$ with $\bfKp$ along the $x$-axis and the line-of-sight direction along the $z$-axis. We have neglected the higher order contributions of $\mathcal O(\phi^2), \mathcal O(\omega^2)$, and $\mathcal O(\phi \omega)$.
Figure~\ref{fig:vecs} shows the relation of the vectors.
Using this, we can obtain the contribution to the connected 4PCF from lensing: 
\begin{align}
     \vev{\delta_{g_1} \delta_{g_2} \delta_{g_3} \delta_{g_4}}_c &= \frac{V^2}{S} \sum_{\bfKp} \delta_{ {\bf k}_{1\perp} +{\bf k_{2\perp},\bfKp}} \delta_{k_{1z},-k_{2z}} \delta_{ {\bf k}_{3\perp} +{\bf k_{4\perp},-\bfKp}} \delta_{k_{3z},-k_{4z}} K_\perp^2\nonumber \\
     & \quad \times \left\{  P_{\phi\phi}(K_\perp)\left( P_1 \cos\theta_1 + P_2 \cos\theta_2 \right) \left( P_3 \cos\theta_3 + P_4 \cos\theta_4\right) \right. \nonumber \\
     & \qquad + P_{\omega\omega}(K_\perp)\left( P_1 \sin\theta_1 + P_2 \sin\theta_2 \right) \left( P_3 \sin\theta_3 + P_4 \sin\theta_4 \right) \nonumber \\
     & \qquad \left. - P_{\phi\omega}(K_\perp) \left[ \left( P_1 \cos\theta_1 + P_2 \cos\theta_2 \right) \left( P_3 \sin\theta_3 + P_4 \sin\theta_4\right) + \left( P_1 \sin\theta_1 + P_2 \sin\theta_2 \right) \left( P_3 \cos\theta_3 + P_4 \cos\theta_4\right) \right] \right\} \nonumber \\
     & \quad + (2 \text{\ other permutations}),
\label{eqn:singlemodefourpoint}
\end{align}
\end{widetext}
where $\delta_{g_i}$ is a shorthand for $\delta_g(\bfk_1)$, $P_i$ for $P_g(k_i) k_{i\perp}$, and we have defined the power spectra
\begin{equation} 
\vev{X(\bfKp) Y^*(\bfK'_\perp)} = S\, \delta_{\mathbf K_\perp, \mathbf K'_\perp} P_{XY}(K_\perp),
\label{eq:xy_corr}
\end{equation} 
with $X,Y \in \{\phi, \omega\}$. 
We have normalized the power spectrum to be consistent with that in the infinite observation area, similar to Eq.~(\ref{eq:ps_3d}).
Note that $P_{\phi\omega} = P_{\omega\phi}$ because $\phi(\mathbf x_\perp)$ and $\omega(\mathbf x_\perp)$ are real fields.
Under a parity inversion, the dot products (the cosines) remain invariant, but the cross products (the sines) change sign. One can then appreciate that the mixed term arising in 
the final line in Eq.~(\ref{eqn:singlemodefourpoint}) is a parity-odd contribution to the 4PCF.\footnote{
Note that Eq.~(\ref{eqn:singlemodefourpoint}), including its final line, is invariant under the change of coordinates from the left-handed ones to right-handed ones because $P_{\omega\phi}$ gives another sign flip. This just means that the 4PCF is a scalar quantity. To see if some terms are parity-odd or even, we must compare two configurations connected through the parity inversion in the same handed coordinates, which changes the sign of the sines without changing the sign of $P_{\phi\omega}$. }

In the above calculation, we have focused solely on the impact of lensing on the angular deflection of galaxies. However, lensing can also lead to area distortions and flux amplification, which are competing effects in the observed galaxy number density \cite{Bartelmann:1994ye,Dodelson:2003bv,Jeong:2012nu}.
This will introduce an overall prefactor to the lensed 4PCF, which will equivalently affect the parity-even and parity-odd components.  

\subsection{The lensing power spectrum from chiral gravitational waves
}
\label{sec:4PCF-GW}

In the above calculation of the 4PCF, we calculated the contribution to the 4PCF from a general lensing deflection. Let us now consider how lensing by chiral GWs leads to a non-zero parity-breaking contribution to the 4PCF in Eq.~\eqref{eqn:singlemodefourpoint}. Recall from above that the parity-breaking contribution in the 4PCF arises from the term proportional to $P_{\phi\omega}(K_\perp)$.
We can obtain this (2D) power spectrum $P_{\phi\omega}(K_\perp)$ from the (3D) chiral GW power spectrum, using results from Refs.~\cite{Book:2011na} and \cite{Dai:2012bc}.  Our small-sky power spectra can be obtained from the full-sky power spectra in those papers by noting the correspondence,
\begin{align}
    \phi(\mathbf x_\perp) = r^2 \Phi(\hat n),\ \   \omega(\mathbf x_\perp) = r^2 \Omega(\hat n), 
    \label{eq:Omega_Phi}
\end{align}
between our potentials $\phi$ and $\omega$ (functions of physical positions whose gradients give physical-distance displacements) and their dimensionless counterparts, $\Phi$ and $\Omega$.  We also use the correspondence $\ell \to r K_\perp$ in the large-$\ell$ limit, and the correspondence $\delta_{\ell\ell'} \delta_{mm'} \leftrightarrow (2\pi)^2 r^{-2}\delta(\bfK_\perp -\bfK_\perp')$.  Our parity-breaking power spectrum is then \cite{Li:2006si,Book:2011na}, 
\begin{align}
        \label{eq:cl_phi_ome}
    P_{\phi\omega}(K_\perp) &= r^6 C_{\ell=rK_\perp}^{\Phi\Omega} \nonumber \\
   &= r^6\int \frac{k^2\, \dd k}{2\pi^2}\, \left[ P_L(k)-P_R(k) \right]F^\Phi_\ell(k)F^\Omega_\ell(k),
\end{align}
where $P_L$ and $P_R$ are the left- and right-handed GW power spectra, respectively, defined by:
\begin{equation} 
\langle h_{R,L}({\bf k})h_{R,L}({\bf k'})^*\rangle = (2\pi)^3 \delta({\bf k - k'})P_{R,L},
\end{equation} 
where $h_{R,L}$ are the right- and left-handed GW modes, respectively,
and\footnote{
The last line in Eq.~(\ref{eq:f_phi}) is missing in Eq.~(33) of Ref.~\cite{Book:2011na}.} 
\begin{align}
    \label{eq:f_ome}
    F^\Omega_\ell(k) &= N_\ell \int^{k\eta_0}_{k\eta_r} \dd w \,T(w) \frac{j_\ell(k\eta_0-w)}{(k\eta_0-w)^2}, \\
    F^\Phi_\ell(k) &= - N_\ell \int^{k \eta_0}_{k\eta_r} \dd w \,T(w) \frac{1}{k(\eta_0-\eta_r)}\nonumber \\
    & \times \left\{(k\eta_0-w)\left[ \frac{\partial}{\partial w} + \frac{1}{2}(w- k\eta_r) \left(1 + \frac{\partial^2}{\partial w^2} \right) \right] \right. \nonumber \\
    &\qquad
    \left. -3 - 2(w - k\eta_r) \frac{\partial}{\partial w} \right\} \frac{j_\ell(k\eta_0 - w)}{(k\eta_0 - w)^2},
    \label{eq:f_phi}
\end{align}
with $\eta_0$ the present conformal time, $\eta_r \equiv \eta_0 -r$, $T(w)=3j_1(w)/w$ in terms of spherical Bessel function of the first kind $j_\ell(w)$, and $N_\ell \equiv \sqrt{2\pi(\ell+2)!/(\ell-2)!}/(\ell(\ell+1))$.
For completeness, the power spectra for the auto-correlations of $\phi$ and $\omega$ are 
\begin{align}
    P_{X  X}(K_\perp) &\simeq \frac{r^6}{2\pi^2} \int k^2\, dk\, \left[ P_L(k) + P_R(k) \right]\left[F^{\bar X}_{\ell}(k)\right]^2,
\end{align}
where $\bar X \in \{\Phi,\Omega\}$. 

We note that, although we have assumed the continuous limit for the 2D Fourier modes so far, we can easily relate it to the finite-arc case with $S = s^2$ as 
\begin{align}
    \vev{X(\bfKp) Y^*(\bfK'_\perp)} &= (2\pi)^2\delta(\bfKp - \bfK'_\perp)P_{XY}(\bfKp) \nonumber \\
    &\simeq S\, \delta_{\bfKp, \bfK'_\perp} P_{XY}(\bfKp).
\end{align}

\section{Detectability}
\label{sec:detectability}
Having obtained an expression for the parity-breaking contribution of the 4PCF from chiral GWs in Eq.~\eqref{eqn:singlemodefourpoint}, we now turn to a discussion of the detectability of this signal.
Following the analysis outlined in Ref.~\cite{Jeong:2012df},
we first obtain the estimator for $\phi$ and $\omega$ with one pair of ($\mathbf k_1,\mathbf k_2)$ from Eq.~(\ref{eq:n_2pt}):

\begin{align}
\widehat{X_{\bfk_1,\bfk_2}(\bfKp)} &= \delta_g(\mathbf k_{1\perp},k_{1z}) \delta_g(\mathbf k_{2\perp},k_{2z}) f_X(\mathbf k_{1\perp}, \mathbf k_{2\perp},k_{1z})^{-1},
\end{align}
where $X \in \{\phi,\omega\}$ again, $k_{1z} = -k_{2z}$, $\bfKp = \bm k_{1\perp} + \bm k_{2\perp}$, and 
\begin{align}
    \label{eq:f_phi_de}
    f_\phi(\mathbf k_{1\perp}, \mathbf k_{2\perp},k_{1z}) &= \frac{V K_\perp}{S} \left[P_1 \cos \theta_{12} + P_2 \cos \theta_{21} \right], \\
    \label{eq:f_ome_de}    
    f_\omega(\mathbf k_{1\perp}, \mathbf k_{2\perp},k_{1z}) &= -\frac{V K_\perp}{S} \left[ P_1 \sin \theta_{12} + P_2 \sin \theta_{21} \right].
\end{align}
Note that, while both $f_\phi$ and $f_\omega$ are symmetric under $\mathbf k_1 \leftrightarrow \mathbf k_2$, $f_\phi$ is symmetric but $f_\omega$ is asymmetric under $\bfk_{1\perp} \leftrightarrow \bfk_{2\perp}$ with $k_{1z}$ and $k_{2z}$ fixed, which is due to the parity odd property of the curl mode ($\omega$) in the 2D celestial space. 
The variances of these estimators are given by\footnote{
We define the variance with the explicit $S$ in the right-hand side of Eq~(\ref{eq:aa_sigma}) so that $\sigma_X^2$ has the same dimension of $P_X(K_\perp)$. Even if we define the $\sigma_X^2$ including $S$, the final results do not change.
}
\begin{align}
    \label{eq:aa_sigma}
    \vev{\widehat {X_{\bfk_1,\bfk_2}(\bfKp)} \widehat {X_{\bfk_1,\bfk_2}(\bfKpp)}^*} &= S\, \delta_{\bfKp, \bfKpp} \sigma_{X,{\bfk_1,\bfk_2}}^2, 
\end{align}
where $\sigma_X^2$ is defined as 
\begin{equation}
    \sigma^2_{X,{\bfk_1,\bfk_2}} = 2 \frac{V^2}{S} P^{\tot}(k_1)P^\tot(k_2) f_X(\mathbf k_{1\perp}, \mathbf k_{2\perp},k_{1z})^{-2},
\end{equation}
with $P^\tot = P_g + P_n$ with $P_n$ the noise power spectrum. We introduce the noise power spectrum to take into account the smallest scale that observations can reach,  which could come from instrumental precisions and/or the non-linear evolution of density perturbations.
The coefficient $2$ in this expression comes from the fact that there are two permutations in Eq.~(\ref{eq:aa_sigma}) for the connected contribution. Also, $V^2$ comes from the definition of the power spectrum (see Eq.~(\ref{eq:ps_3d})).

By summing over all pairs of $(\mathbf k_1, \mathbf k_2)$ with inverse-variance weighting, we obtain the minimum-variance estimator for $X(\bfKp)$:
\begin{widetext}
\begin{align}
\widehat {X(\bfKp)} &= P^n_X(\bfKp)\sum_{\mathbf k_\perp} \sum_{k_z}\frac{f_X(\mathbf k_{\perp}, \mathbf K_\perp- \mathbf k_{\perp},k_{z}) \delta_g(\mathbf k_\perp,k_z)  \delta_g(\mathbf K_\perp - \mathbf k_\perp,k_z)}{2(V^2/S) P^{\tot}(k)P^\tot(\sqrt{|\mathbf K_\perp- \mathbf k_\perp|^2 + k_z^2})},
\label{eq:a_est}
\end{align}
where the noise power spectra are given by 
\begin{align}
  P^n_X({\bf K_\perp)} &= \left[\sum_{\mathbf k_\perp} \sum_{k_z}\frac{f_X(\mathbf k_{\perp}, \mathbf K_\perp- \mathbf k_{\perp},k_{z})^2}{2(V^2/S) P^{\tot}(k)P^\tot(\sqrt{|\mathbf K_\perp- \mathbf k_\perp|^2 + k_z^2})} \right]^{-1}.
  \label{eq:p_n}
\end{align}
\end{widetext}

To discuss the detectability, let us here consider the maximally chiral scenario in which the GW background is entirely left-handed, namely $P_R =0$. We can characterize the left-handed GWs as $P_L(k) = A_{L}P^f_{L}(k)$, where $A_{L}$ is the amplitude and $P_{L}^f$ some fiducial power spectrum normalized as $(2\pi^2)^{-1}\int\, k^2 \,\dd k\, P^f_L(k) = 1$.
We here define \begin{align}
        \mathcal M(K_\perp,r) \equiv r^6 \int \frac{k^2 \dd k}{2\pi^2} P^f_{L}(k)F^\Phi_{r K_\perp}(k)F^\Omega_{r K_\perp}(k).
        \label{eq:mk_int}
\end{align}
Then, from Eq.~(\ref{eq:cl_phi_ome}), we can make the estimator for $A_{L}$ with $\bfKp$ mode as 
\begin{align}
    \widehat{A^{r,\bfKp}_{L}} = \left[\mathcal M(K_\perp,r) \right]^{-1}\left[ S^{-1} \widehat {\phi(\bfKp)} \widehat {\omega(\bfKp)}^* \right].
\end{align}
The variance of this estimator is given by
\begin{align}
    \vev{\widehat {A^{r,\bfKp}_{L}} \widehat {A^{r,\bfKp}_{L}}^*} &= \delta_{\bfKp, \bfKpp} \sigma_{A^{r,\bfKp}_{L}}^2, \\
    \sigma^2_{A^{r,\bfKp}_{L}} &= \left[\mathcal M(K_\perp,r) \right]^{-2} P^n_{\phi}(K_\perp) P^n_{\omega}(K_\perp).
\end{align}
By summing over all $\bfKp$ with inverse-variance weighting, we obtain the minimum-variance estimator for $A^r_L$:
\begin{align}
    \widehat{A^r_{L}} = \sigma^{2}_{A^r_{L}} \sum_\bfKp \frac{\left[\mathcal M(K_\perp,r) \right]^{2}}{P^n_{\phi}(K_\perp) P^n_{\omega}(K_\perp)}\left[ S^{-1} \widehat {\phi(\bfKp)} \widehat {\omega(\bfKp)}^* \right],
\end{align}
where 
\begin{align}
    \sigma^{-2}_{A^r_{L}} = \sum_{\bfKp} \frac{\left[\mathcal M(K_\perp,r) \right]^2}{P^n_{\phi}(\bfK_\perp) P^n_{\omega}(\bfK_\perp)},
    \label{eq:sig_a_hl}
\end{align}
is the variance with which $A_L^r$ can be measured. Throughout this paper, we take into account the finite number of the modes satisfying $k_\tmin < \{k_\perp, k_z, K_\perp \} < k_\tmax$, where $k_\tmin$ is determined by the observation box size $k_\tmin \sim 1/s$ and $k_\tmax$ is determined by the smallest scale that the observation can reach.

Let us estimate $\sigma_{A^r_L}$ in some realistic situations.
For simplicity, we assume $P_g/P^\tot = 1$ for $k< k_\tmax$ and $P_g/P^\tot =0$ for $k> k_\tmax$, similar to Ref.~\cite{Jeong:2012df}.
From this assumption and Eq.~(\ref{eq:p_n}), we obtain 
\begin{align}
    P^n_X(\bfK_\perp) \sim \left[\sum^{k_\tmax}_{\mathbf k_\perp} \sum^{k_\tmax}_{k_z} S^{-1} K_\perp^2 k_\perp^2 \right]^{-1} \sim \frac{k_\tmin}{K_\perp^2 k_\tmax^5},
\end{align}
where we have used $s \sim 1/k_\tmin$ and the fact that the dominant contribution comes from the squeezed limit configuration, $K_\perp \ll k_{1\perp} \simeq k_{2\perp}$.\footnote{
The estimator in this work is sensitive to the squeezed configuration. If we input the expected configuration of the signals in specific models, we could optimize the estimator. We leave the analysis for specific models for future work. }
 Note that the number of the modes with $k_\perp \sim \mathcal O(k_\tmax)$ is $\mathcal O(k_\tmax/k_\tmin)$.
We have also taken the angular average for the square of the trigonometric functions from Eqs.~(\ref{eq:f_phi_de}) and (\ref{eq:f_ome_de}), which leads to an $\mathcal O(1)$ factor.

\begin{figure}
        \centering \includegraphics[width=\columnwidth]{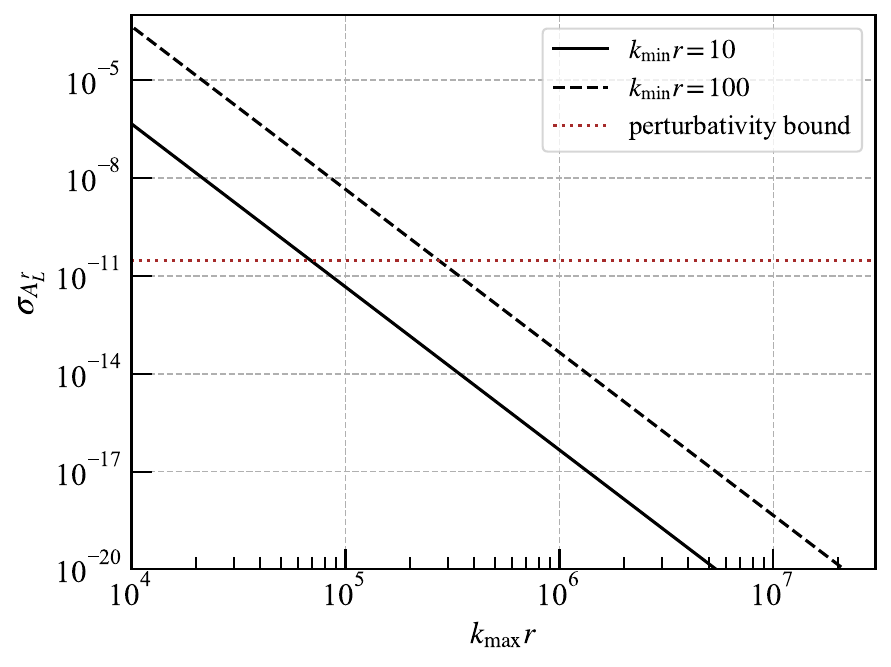}
        \caption{ 
        The order estimate of $\sigma_{A_L^r}$, which corresponds to the minimum $A_{L}$ detectable at $\mathcal O(1)\sigma$ level.
        The black lines are from the right-hand-side of Eq.~(\ref{eq:sigma_ahl}).
        $k_* \eta_0 = 10^4$ is taken for both the lines. 
        $k_\tmin r = 10$ is taken for the black solid line and $k_\tmin r = 100$ for the black dashed line. 
        The brown dotted line shows the upper bound on $A_{L}$ from the perturbativity, Eq.~(\ref{eq:A_up_b}), with $k_* \eta_0 = 10^4$. 
        }
        \label{fig:sigma}
\end{figure}

To obtain the order of $|\mathcal M(K_\perp,r)|$, let us use Limber's approximation~\cite{1953ApJ...117..134L} for the integral of Eq.~(\ref{eq:mk_int}) (see also Ref.~\cite{Breysse:2014uia} for a similar usage of Limber's approximation).
We here assume that $\ell \gg 1$ and the power spectrum, $P^f_{L},(k)$ and the functions other than $j_\ell$ in $\mathcal M$ are slowly varying functions of $k$ compared to the oscillatory $k$-dependence of $F^\Omega$ and $F^\Phi$, which comes from the spherical Bessel function, $j_\ell$.
Strictly speaking, $T(w)$ cannot be considered as a slowly varying function of $k$ because of the oscillatory $k$-dependence. 
However, we can expect that its $k$-dependence does not change the order of our estimate on $\mathcal M(K_\perp,r)$ because both $j_\ell$ and $T$ have the same oscillatory $k$-dependence as $\sin/\cos(k\eta)$.
Specifically, we use the following relation, which underlies Limber's approximation for large $\ell$:
\begin{align}
    \frac{2}{\pi} \int^\infty_0 k^2 \dd k j_\ell(k(\eta_0-\eta)) j_\ell(k(\eta_0-\eta')) = \frac{1}{(\eta_0-\eta)^2} \delta(\eta - \eta').
    \label{eq:limber}
\end{align}
Then, we can reexpress Eq.~(\ref{eq:mk_int}) as 
\begin{widetext}
\begin{align}
    \mathcal M(K_\perp,r) &\sim \left.-r^6 \frac{N_\ell^2}{8\pi} \int^{\eta_0}_{\eta_r} \dd \eta\, k^2 P^f_{L}(k) \frac{1}{(\eta_0-\eta)^2} T^2(k\eta) \frac{\ell^2 (\eta - \eta_r)}{k^5 (\eta_0-\eta_r)(\eta_0 - \eta)^5}\right|_{k = \frac{\ell}{\eta_0-\eta}, \ell = rK_\perp} \nonumber \\
    &\sim \left.-r^6 \frac{N_\ell^2}{8\pi}  \int^{\eta_0-\eta_r}_{0} \dd \bar \eta\, \frac{\ell^2}{\bar \eta^4} P^f_{L}\left( \frac{\ell}{\bar \eta} \right) T^2\left(\frac{\ell (\eta_0-\bar \eta)}{\bar \eta} \right) \frac{(\eta_0 - \bar \eta - \eta_r)}{\ell^3 (\eta_0-\eta_r)}\right|_{\ell = rK_\perp},    
    \label{eq:mk_approx}
\end{align}
where $\bar \eta = \eta_0 - \eta$ and we have used $(1+\partial^2/\partial w^2)j_\ell(k\eta_0 - w) \simeq \ell^2 j_\ell(k\eta_0 - w)/(k\eta_0-w)^2$ and we have neglected the other terms in Eq.~(\ref{eq:f_phi}) because they are subdominant in $\ell \gg 1$. 

We here assume that the GW power spectrum has a large peak around $k_*$ with a width of $\Delta k \simeq \mathcal O(1) k_*$. 
Then, we can rewrite Eq.~(\ref{eq:mk_approx}) as 
\begin{align}
   \mathcal M(K_\perp,r) 
    &\sim \left.-r^6 \frac{\pi}{4} N_\ell^2 \int^{\infty}_{\ell/(\eta_0-\eta_r)} \frac{\bar k^2 \dd \bar k}{2\pi^2} \frac{1}{\ell} P^f_{L}( \bar k ) T^2\left( \bar k \eta_0 - \ell \right) \frac{(\eta_0 - \ell/\bar k - \eta_r)}{\ell^3 (\eta_0-\eta_r)}\right|_{\ell = rK_\perp} \nonumber \\   
    &\sim \left.-r^6 \frac{\pi}{4} N_\ell^2 T^2\left( k_* \eta_0 - \ell \right) \frac{(k_*\eta_0 - k_*\eta_r - \ell)}{\ell^4 k_*(\eta_0-\eta_r)}\right|_{\ell = rK_\perp} \nonumber \\
    &\sim \left.-r^6 \frac{9\pi}{8} N_\ell^2 (k_*\eta_0)^{-4} \ell^{-4}\right|_{\ell = rK_\perp},
\end{align}
\end{widetext}
where $\bar k \equiv \ell/\bar \eta$ and we have assumed that $k_*(\eta_0 - \eta_r) \gg \ell$ and approximated $T^2(x) \simeq 9/(2x^4)$ in $x \gg 1$ by taking the oscillation average because we are interested in the order of $\mathcal M$.
Substituting this into Eq.~(\ref{eq:sig_a_hl}), we obtain 
\begin{align}
     &\sigma^{-2}_{A^r_{L}} \sim  \sum_{\bfKp} \left( \mathcal M(K_\perp,r) \frac{K_\perp^2 k_\tmax^5}{k_\tmin}\right)^2 \nonumber \\
     & \quad \ \ \, \sim \left( \mathcal M(k_\tmin,r) k_\tmin k_\tmax^5\right)^2 \nonumber \\
     \Rightarrow \quad
     &\sigma_{A^r_{L}} \sim \left( |\mathcal M(k_\tmin,r)| k_\tmin k_\tmax^5\right)^{-1},
     \label{eq:sigma_ahl}
\end{align}
where we have used that the summation over $\bfKp$ is dominated by the smallest $K_\perp (\sim k_\tmin)$ because of the negative power of $\ell (=r K_\perp)$ in $\mathcal M$.
To detect the signal with $\mathcal O(1) \sigma$ level, $A_{L} > \sigma_{A^r_L}$ must be satisfied. 
Figure~\ref{fig:sigma} shows the $k_\tmax r$ dependence of $\sigma_{A^r_L}$ with some fiducial values.

On the large scales ($k\lesssim 0.1\,\text{Mpc}^{-1}$), GWs are constrained from the CMB B-mode observations. 
Also, on the small scales ($k >1\,\text{Mpc}^{-1}$), GWs are constrained from the observational upper bound on the degrees of freedom of relativistic particles.
As an interesting example, we here consider the case where the GW spectrum has a peak on the intermediate scale, $k_*\eta_0 \sim 10^4$ with $\eta_0 \simeq 14\, \text{Gpc}$ ($k_* \sim \mathcal O(0.1)\,\text{Mpc}^{-1}$), where GWs are not constrained by the above two~\cite{Clarke:2020bil}.
However, we impose that they are in the perturbative regime before the horizon crossing, $k_*^3/(2\pi^2)P_L(k_*)|_{\eta < 1/k_*} < 1$.
After the horizon crossing, the tensor perturbations get red-shifted and finally determine 
 the current amplitude of the power spectrum, $A_L$.
From this, we obtain the upper bound on the intermediate scales for $A_{L}$:
\begin{align}
A_{L} < 3\times 10^{-11} \left(\frac{k_*\eta_0}{10^4}\right)^{-2} \  (0.1 \lesssim k_*/\Mpc^{-1} \lesssim 1),
\label{eq:A_up_b}
\end{align}
where we have used that the redshift at the horizon crossing during the radiation era, $aH (=2/\eta) = k_* = 10^4/\eta_0$, is $z \simeq 1.7\times 10^5$ and approximated that $P_L$ evolves as $\propto 1/a^2$ where $a$ is the scale factor after the horizon crossing.

Let us finally consider the most promising case where the upper bound, Eq.~(\ref{eq:A_up_b}), is saturated. 
If we take $k_\tmin r = 10$ and $k_*\eta_0 = 10^4$, we find $k_\tmax r \gtrsim \mathcal O(10^4)$ required from Fig.~\ref{fig:sigma}. 
For example, if we further assume $\eta_r/\eta_0 =0.5$ (corresponding to the redshift $z \simeq 4$ at $r$), the necessary condition for the parity-violation detection is $k_\tmax \gtrsim \mathcal O(1)\,\text{Mpc}^{-1}$ with $k_\tmin = 1.43\times 10^{-3}\,\text{Mpc}^{-1}$.
 While this range may not be feasible for current galaxy surveys, it could be within the reach of future 21 cm or other line intensity mapping observations \cite{Kovetz:2017agg}.

\section{Chiral Gravitational-wave background models}
\label{sec:models}

In the above calculation, we remained agnostic to the source of the chiral GW background which induces the parity-breaking 4PCF. In this Section, we turn to a brief discussion of models and scenarios that could give rise to such a background. In order for the parity-breaking contribution to the galaxy 4PCF to be non-zero, the GW background must be chiral, such that $P_R \neq P_L$, arising from some parity violation in the gravitational sector. While GR itself is a parity invariant theory and predicts GWs as such, there are a wide range of extensions to GR as well as early-universe scenarios that produce the necessary chiral GW background to source the 4PCF. 

One well-studied example is the addition of a gravitational Chern-Simons term \cite{Lue:1998mq, Jackiw:2003pm, Alexander:2009tp}: 
\begin{equation}
    \mathcal{L} \supset \alpha\varphi R\tilde{R},
\end{equation}
where $\varphi$ can either be the inflaton or an auxiliary pseudoscalar, $\alpha$ is a coupling constant, and $R\tilde{R}$ is the Pontryagin density of the spacetime, defined as 
\begin{equation}
    R \tilde{R} = \frac{1}{2}\epsilon^{\rho\sigma\alpha\beta}R^\mu{}_{\nu\alpha\beta}R^\nu{}_{\mu\rho\sigma}.
\end{equation}
Because $R\tilde{R}$ is parity-violating and $\varphi$ is a \textit{pseudo}scalar, the propagation of GWs in Chern-Simons gravity will be chiral, such that one of the right- or left-handed polarizations will be amplified and the other will be attenuated \cite{Lue:1998mq, Alexander:2007kv}. This effect has been well studied in the case of GWs from late universe compact binaries \cite{Yunes:2010yf,Alexander:2017jmt, Yagi:2017zhb, Zhao:2019xmm, Qiao:2019wsh, Okounkova:2021xjv, Callister:2023tws, Ng:2023jjt,Daniel:2024lev,Lagos:2024boe}, but also applies to the propagation of inflationary GWs that source our 4PCF, as discussed in e.g. \cite{Alexander:2004wk, Bartolo:2017szm, Nojiri:2020pqr, Cai:2022lec}.\footnote{Ref. \cite{Creque-Sarbinowski:2023wmb} has also shown that the addition of a gravitational Chern-Simons term during inflation will lead to a parity violating 4PCF due to particle exchange. }  In addition to Chern-Simons gravity, there are other modified gravity models containing parity-violating curvature invariants, which can also lead to chiral GWs. See Ref.~\cite{Jenks:2023pmk} for an overview of different models and modifications to the GW waveforms. Schematically, these GWs are modified from GR as 
\begin{equation} 
h_{R,L} = h_{R,L}^{\rm GR}e^{\mp \kappa(\varphi, f, z)},
\end{equation} 
where $h_{R,L}^{\rm GR}$ is the unmodified, GR expression for $h_{R,L}$ and $\kappa$ is a function which depends on the scalar, $\varphi$, as well as the frequency, $f$, and propagation distance, $z$, of the GWs. We can clearly see that a modification to the GWs of this form will lead to the desired $P_{R} - P_{L} \neq 0$ required to induce a parity-breaking 4PCF.

There are also models in which inflationary chiral GWs arise from interactions with an axion and gauge fields~\cite{Sorbo:2011rz, Anber:2012du, Adshead:2013qp, McDonough:2018xzh,Bastero-Gil:2022fme}. In these scenarios, the Lagrangian includes a term of the form 
\begin{equation}
    \mathcal{L} \supset \frac{\alpha \varphi}{4 f_a} F_{\mu\nu}\tilde{F}^{\mu\nu},
\end{equation}
where as above, $\varphi$ can either be the inflaton or an additional pseudoscalar, $\alpha$ is a coupling constant, $f_a$ is the axion decay constant, and $F_{\mu\nu} = \partial_\mu A_\nu - \partial_\mu A_\nu$ is the field strength of a gauge field, $A_\mu$. In these scenarios, a chiral GW background is sourced by the gauge fields and will also lead to a parity-breaking galaxy 4PCF.
At the same time, these models also predict the parity breaking in the 4PCF of curvature perturbations through the exchange of the gauge particles during inflation (see e.g., \cite{Fujita:2023inz,Niu:2022fki}). 

We leave a detailed analysis of the effects of specific models on the 4PCF to future work. However, we emphasize that early-universe scenarios that generate chiral GWs are well-motivated and numerous. Thus, the general 4PCF signature calculated in this work is well posed from a wide range of inflationary scenarios.

 \section{Conclusions}\label{sec:conclusion}
Motivated by the recent hints of parity violation in cosmological data, in this paper we have considered the possibility that lensing by a chiral GW background can lead to a parity-breaking galaxy 4PCF. We showed that a general lensing deflection does indeed lead to a parity-breaking contribution to the 4PCF. This contribution will be nonzero if the GW background doing the lensing is chiral, $P_R \neq P_L$. Since this kind of parity-breaking signals come from the lensing effects on the (approximate) 2D observation plane of galaxies, they are physically different from the parity breaking signals reported in Refs.~\cite{Hou:2022wfj,Philcox:2022hkh}, which are based on the observation of the tetrahedron configurations of the galaxies.
That said, we stress that our system is 3D because it includes the observer in addition to the 2D observation plane.

We then estimated the feasibility of detecting such a signal with current or forthcoming galaxy surveys and found that it may be within the reach of future 21cm observations. Lastly, we commented on parity-breaking GW models that may lead to the production of the necessary chiral GWs. 

While the lensing that induces parity-violating 4PCF, discussed in this paper, may not be responsible for the current hint of parity violation in the large-scale structure data, it is worth forecasting in more detail how such an effect could be discerned with future observations. We also note that if a parity-breaking 4PCF is observed and thought to arise from lensing, one could check this using cosmic shear estimators from galaxy shape distortions. Furthermore, having shown this proof of principle for a generic parity-violating GW background, it would be interesting to determine the parity-violating 4PCF signatures for particular models such as those discussed in Section~\ref{sec:models} and others. We defer these studies and others to future work.

\section*{Acknowledgements}

KI is supported by JSPS Postdoctoral Fellowships for Research Abroad.
LJ is supported by the Kavli Institute for Cosmological Physics at the University of Chicago.
This work was supported at Johns Hopkins by NSF Grant No.\ 2112699, the Simons Foundation, and the Templeton Foundation.

\appendix

\begin{widetext}
    \section{Derivation of Eq.~(\ref{eq:delta_g_fourier})}

In this appendix, we show the derivation of Eq.~(\ref{eq:delta_g_fourier}).
The Fourier series of $\delta_g$ are given by
\begin{align}
    \label{eq:expa_3d}
    \delta_g(\bfx) &= \frac{1}{V}\sum_{\bfk} \delta_g(\bfk) \ee^{i \bfk \cdot \bfx}, \ 
    \delta_g(\bfk) = \int_{V}\dd^3 x\, \delta_g(\bfx) \ee^{-i \bfk \cdot \bfx}.
\end{align}
Similarly, we expand a 2D field $X(\bfx_\perp)$ as 
\begin{align}
    X(\bfx_\perp) &= \frac{1}{S}\sum_{\bfK_\perp} X(\bfK_\perp) \ee^{i \bfK_\perp \cdot \bfx_\perp}, \ 
    X(\bfK_\perp) = \int_{S}\dd^2 x\, X(\bfK_\perp) \ee^{-i \bfK_\perp \cdot \bfx_\perp}.  
\end{align}
Substituting Eq.~(\ref{eq:delta_g_x}) into Eq.~(\ref{eq:expa_3d}), we reproduce Eq.~(\ref{eq:delta_g_fourier}):
\begin{align}
    \delta_g(\bfk) &= \int_{V}\dd^3 x\, \left\{\delta_{g,0}(\bfx) + [\partial_i \phi(\bfx_\perp) + \epsilon_{ij} \partial_j \omega(\bfx_\perp)] \partial_i \delta_{g,0}(\bfx) \right\} \ee^{-i \bfk \cdot \bfx} \nonumber \\
    &= \delta_{g,0}(\bfk) - \int_{V}\dd^3 x\, \frac{1}{SV}\sum_{\bfk'} \sum_{\bfK_\perp} [\bfK_i \phi(\bfK_\perp) + \epsilon_{ij} \bfK_j \omega(\bfK_\perp)] \bfk'_i \delta_{g,0}(\bfk') \ee^{-i (\bfk - \bfk'- \bfK_\perp) \cdot \bfx} \nonumber \\
    &= \delta_{g,0}(\bfk_\perp, k_z) + S^{-1}\sum_{\bfKp} \left[ - \bfKp\cdot(\bfkp-\bfKp) \phi(\bfKp)  + \bfKp\times(\bfkp-\bfKp) \omega(\bfKp)\right] \delta_{g,0}(\bfkp-\bfKp,k_z).
\end{align}

\end{widetext}

\bibliography{draft_pb_4pcf}

\end{document}